\documentstyle[12pt,bezier]{article}
\begin{document}
\def \inbar{\vrule height1.5ex width.4pt depth0pt}
\def \C{\relax\hbox{\kern.25em$\inbar\kern-.3em{\rm C}$}}
\def \R{\relax{\rm I\kern-.18em R}}
\newcommand{\Z}{\ Z \hspace{-.08in}Z}
\newcommand{\be}{\begin{equation}}
\newcommand{\ee}{\end{equation}}
\newcommand{\bea}{\begin{eqnarray}}
\newcommand{\eea}{\end{eqnarray}}
\newcommand{\nn}{\nonumber}
\newcommand{\kt}{\rangle}
\newcommand{\br}{\langle}
\newcommand{\lll}{\left( }
\newcommand{\rrr}{\right)}
\newcommand{\Psusy}{Parasupersymmetric}
\newcommand{\psusy}{parasupersymmetric}
\newcommand{\tii}{topological invariant}
\newcommand{\tiis}{topological invariants}
\newcommand{\Tii}{Topological invariant}
\newcommand{\qq}{{\cal Q}}
\newcommand{\ddd}{\Delta^{(p=2)}}
\newcommand{\psqm}{($p=2$)--PSQM}

\title{ Parasupersymmetric Quantum Mechanics  ~~~~~~~~~ ~~~~~
{}~ and  Indices of Fredholm Operators}
\author{Ali Mostafazadeh\thanks{E-mail: alimos@phys.ualberta.ca}\\ \\
Theoretical Physics Institute,
 University of Alberta, \\
Edmonton, Alberta,
 Canada T6G 2J1.}
\date{ June 1996}
\maketitle

\begin{abstract}
The general features of the degeneracy structure of ($p=2$) parasupersymmetric
quantum mechanics are employed to yield a classification scheme for the form of
the parasupersymmetric Hamiltonians. The method is applied to
parasupersymmetric systems whose Hamiltonian is  the square root of a forth
order polynomial in the generators of the parasupersymmetry. These systems are
interesting to study for they lead to the introduction of a set of topological
invariants very similar to the Witten indices of ordinary supersymmetric
quantum mechanics.  The topological invariants associated with
parasupersymmetry are shown to be related to a pair of Fredholm operators
satisfying two compatibility conditions. An explicit algebraic expression for
the topological invariants of a class of parasupersymmetric systems 
is provided.
\end{abstract}

\newpage 

\baselineskip=18pt

\section{Introduction}
Perhaps one of the most intriguing aspects of supersymmetry is its relation
with the Atiyah-Singer index theorem \cite{as}. It was Witten \cite{w1} who
first recognized this relation in the context of  supersymmetric quantum
mechanics (SQM). The subsequent developments in this direction have led to
supersymmetric proofs of the index theorem \cite{a-g-w}.

Few years after the publication of the first supersymmetric proofs of the index
theorem Rubakov and Spiridonov (R-S) \cite{rs}  introduced their
($p=2$)-parasupersymmetric quantum mechanics (PSQM). This  involved a
generalization of the superalgebra of SQM\footnote{Here $\kappa$ is a
conventional positive constant.}:
	\be
	 {\cal Q}^2= [  {\cal Q}, H ] \: =\: 0\,, ~~
	\{ {\cal Q} , {\cal Q}^\dagger \} = 2\kappa \, H\;,  ~~~  (\kappa\in\R^+)
	\label{q1}
	\ee
namely the {\em parasuperalgebra}:
        \bea
        \qq^3\:=\:0\;,~\; [\qq ,H]&=& 0 \label{q0.1}\;, \\
        \{\qq^2,\qq^\dagger\}+\qq\qq^\dagger\qq&=&4\qq H\;.
        \label{q0.2}
        \eea
The latter relations (\ref{q0.1}) and (\ref{q0.2}) have since been generalized
to arbitrary order $p>2$, by Khare \cite{kh1}, and modified by Beckers and
Debergh (B-D) \cite{bd}. B-D ($p=2$)-parasuperalgebra is given by
Eqs.~(\ref{q0.1}) and
        \be
        \left[ \qq ,\left[\qq^\dagger, \qq\right]\right]=2\qq H\;.
        \label{q0.4}
        \ee

It is not difficult to check that indeed superalgebra of SQM (\ref{q1}),  with
$\kappa=2$ and $\kappa=1/2$, is a special case of the parasuperalgebras of R-S
and B-D, respectively. Given the relation between SQM and topological
invariants such as indices of Fredholm operators, the latter observation
renders the following questions:
	\begin{itemize}
	\item[1) ] {\em Is PSQM related to topological invariants? } If so,
	\item[2) ] {\em Are these invariants more general than the indices of
Fredholm operators?}
	\end{itemize}

In a preceding article \cite{p8}, it is shown that a careful analysis of  the
defining parasuperalgebra (for both R-S and B-D types) provides important
information on the degeneracy structure of the spectrum of  the corresponding
systems. In particular, postulating the existence of a parasupersymmetry
involution (chirality) operator and supplementing (either of) the
parasuperalgebra(s) with an additional relation expressing the Hamiltonian in
terms of the parasupercharges, namely\footnote{The Hamiltonian (\ref{q0.41})
was originally proposed by Khare et al. \cite{kh2} for different purposes.}
        \be
        H=\frac{1}{2}\left[
        (\qq\qq^\dagger)^2+(\qq^\dagger\qq)^2-
        \frac{1}{2}(\qq\qq^{\dagger 2}\qq+\qq^\dagger \qq^2
        \qq^\dagger)\right]^\frac{1}{2}\;,
	\label{q0.41}
        \ee
one can show that the integer
        \bea
        \Delta^{(p=2)}&:=&n^{\pi B} - 2n^{\pi F}\: =\:
        n^{\pi B}_0 -2n^{\pi F}_0\;, ~~~~~~{\rm with} \label{q0.5}\\
        n^{\pi B}&:=&\mbox{number of parabosonic states}\nn\\
        n^{\pi F}&:=&\mbox{number of parafermionic states}\nn \\
        n^{\pi B}_0&:=&\mbox{number of zero energy parabosonic states}\nn\\
        n^{\pi F}_0&:=&\mbox{number of zero energy parafermionic states}\nn
        \eea
is a topological invariant. Furthermore, it is shown in \cite{p8} that
$\Delta^{(p=2)}$ is a measure of parasupersymmetry breaking, i.e., the
condition $\Delta^{(p=2)}\neq 0$ implies the exactness of parasupersymmetry. In
this respect, it is quite similar to the Witten index of supersymmetry.
However, unlike the Witten index a mathematical interpretation of this
invariant has been lacking until now.

A major difference between SQM and PSQM is that unlike the defining
(super)algebra (\ref{q1}) of SQM, the defining (parasuper)algebras
(\ref{q0.1}), (\ref{q0.2}) and (\ref{q0.1}), (\ref{q0.4}) do not provide an
expression for the Hamiltonian. In fact, one can easily see that the
form of the Hamiltonian in terms of the generators in both R-S and B-D
PSQM is not unique. Therefore, a relevant problem is to seek a
classification scheme for all possible forms of the PSQM Hamiltonian. 

The purpose of the present article is twofold. First it is shown that the
developments reported in Ref.~\cite{p8} can be used, with some additional
effort, to devise a classification scheme for the PSQM Hamiltonians. This
scheme is then used  to study the topological content of ($p=2$) PSQM 
in a broader framework and to provide a precise mathematical interpretation
for the corresponding topological invariants.

In Sec.~2, a brief discussion of SQM
is presented to demonstrate the motivation for the proceeding analysis of
PSQM.  Sec.~3 summarizes  the necessary back ground
information on  the degeneracy structure of the ($p=2$) PSQM. This information
is then used to address the classification problem. In particular,  the moduli
space of all ($p=2$) PSQM systems admitting a Hamiltonian whose square
is a forth order polynomial in the generators of parasupersymmetry, is
constructed.  Here the subspace of systems for which
$\Delta^{(p=2)}$ is a topological invariant is identified. In Sec.~4,   these
invariants are shown to be related to the indices of a pair of Fredholm
operators, thus providing the mathematical interpretation of interest. Sec.~5
includes the concluding remarks.

\section{SQM and the Index Theorem}
The main ingredient of SQM which makes its relation with the index theory
possible, is its simple degeneracy structure. More precisely, the degeneracy
structure of the spectrum of any supersymmetric quantum mechanical system is
determined using only the defining superalgebra (\ref{q1}) and the properties
of the supersymmetry involution (chirality) operator  $\tau$:
        \be
        \tau^2=1,\; \tau^\dagger=\tau,\; \{\qq,\tau\}=0\;.
        \label{q2}
        \ee
In Eqs.~(\ref{q1}) and~(\ref{q2}), $\qq$ stands for (one of) the generator(s)
of supersymmetry, $\qq^\dagger$ is its adjoint, and $H$ is the Hamiltonian. The
chirality operator $\tau$ induces a double grading of the Hilbert space, ${\cal
H}={\cal H}_+\oplus{\cal H}_-$, where
        \be
        {\cal H}_\pm :=\left\{ \psi\in{\cal H}\: :\: \tau\psi=\pm\psi
        \right\}\;.\label{q3}
        \ee
The superalgebra (\ref{q1}) can be employed to show that the energy spectrum is
non-negative and that each positive energy state of definite chirality is
accompanied with another state of the same energy and opposite chirality,
\cite{w1,p8}. In this sense, one says that the positive energy levels are {\em
doubly degenerate}.

Introducing the self-adjoint generators:
        \be
        Q_1=\frac{1}{\sqrt{2}}(\qq+\qq^\dagger)\;,~~~~
        Q_2=\frac{-i}{\sqrt{2}}(\qq-\qq^\dagger)\;,
        \label{q4}
        \ee
one rewrites the superalgebra (\ref{q1}) in the form:
        \bea
        \{ Q_1,Q_2\}&=& 0\;,\label{q1.1}\\
        Q_1^2\:=\: Q_2^2&=&H\;,\label{q1.2}\\
        \left[ Q_1,H\right]&=&0\;,\label{q1.3}\\
        \left[ Q_2,H\right]&=&0\;,\label{q1.4}\\
        \{Q_1,\tau\}&=&0\;,\label{q1.5}\\
        \{Q_2,\tau\}&=&0\;,\label{q1.6}\\
        \tau^2\:=\:1&,&\;\tau^\dagger\:=\:\tau\;.\label{q1.7}
        \eea
In view of (\ref{q1.3}), one can use the eigenvalues $E$ and $q_1=\pm\sqrt{E}$
of $H$ and $Q_1$, to label the states. Here we choose not to include any other
quantum numbers. Their presence will not interfere with the arguments presented
in this article.

For each positive energy level $(E>0)$,the $\{ |E,\pm\sqrt{E}\kt\}$ basis may
be used to yield matrix representations of the relevant operators \cite{p8}.
Alternatively, one may adapt a basis in which $H$ and $\tau$ are diagonal. In
such a basis one has:
        \bea
        \left. Q_1\right|_{{\cal H}_E}&=&\sqrt{\kappa E}\left(
\begin{array}{cc}
        0&-i\\
        i&0\end{array}\right)=\sqrt{\kappa E}\sigma_2\;,~~~~~~~~~~~~~~~~\nn\\
        \left. Q_2\right|_{{\cal H}_E}&=&\sqrt{\kappa E}\left(
\begin{array}{cc}
        0&1\\
        1&0\end{array}\right)=\sqrt{\kappa E}\sigma_1,\label{q5}
{}~~~~~~~~~~~~~~~~\\
        \left. \tau\right|_{{\cal H}_E}=\left( \begin{array}{cc}
        1&0\\
        0&-1\end{array}\right)&=&\sigma_3,~~~~~~~
        \left. H\right|_{{\cal H}_E}=E\left( \begin{array}{cc}
        1&0\\
        0&1\end{array}\right)\;.\nn
        \eea
where ${\cal H}_E$ denotes the eigenspace associated with the eigenvalue $E$
and $\sigma_i$, $i=1,2,3$, are Pauli matrices. The fact that $trace(\left.
\tau\right|_{{\cal H}_E})=0$ is the very reason for the topological invariance
of the Witten index \cite{w1}:
        \bea
        {\rm index}_W&:=&{\rm trace}(\tau)\:=\:
        n^B-n^F\:=\:n^B_0-n^F_0\;, \label{q6}\\
        n^{B}&:=&\mbox{number of bosonic states}\nn\\
        n^{F}&:=&\mbox{number of fermionic states}\nn \\
        n^{B}_0&:=&\mbox{number of zero energy bosonic states}\nn\\
        n^{F}_0&:=&\mbox{number of zero energy fermionic states}\nn
        \eea

Eq.~(\ref{q5}) serves as a motivation for relating the Witten index with the
analytic indices of Fredholm operators. To demonstrate this relationship, first
one introduces the representation
        \be
        {\cal H}=\left(\begin{array}{c}
        {\cal H}_+\\
        {\cal H}_-\end{array}\right)
        \;\label{q7}
        \ee
of the Hilbert space in which $\tau$ is (block-)diagonal. To obtain  the
representations of $Q_i$ ($i=1,2$), one appeals to Eqs.~(\ref{q1.5}) and
(\ref{q1.6}). These together with (\ref{q5}) suggest:
        \be
        Q_1=\left(\begin{array}{cc}
        0&-iD_1^\dagger\\
        iD_1&0\end{array}\right)\;,\;
        Q_2=\left(\begin{array}{cc}
        0&D_2^\dagger\\
        D_2&0\end{array}\right)\;,
        \label{q8}
        \ee
where $D_i:{\cal H}_+\to{\cal H}_-$, $i=1,2$ are a couple of operators acting
on ${\cal H}_+$ and $D_i^\dagger$ are their adjoints. Enforcing the
superalgebra, namely Eqs.~(\ref{q1.1}) and (\ref{q1.2}), this representation
leads to the following set of compatibility conditions  for $D_i$:
        \bea
        D_1^\dagger D_2&=&D_2^\dagger D_1 \;,\label{q9}\\
        D_1 D_2^\dagger &=&D_2 D_1^\dagger \;,\label{q10}\\
        D_1^\dagger D_1&=&D_2^\dagger D_2 \;,\label{q11}\\
        D_1 D_1^\dagger &=&D_2 D_2^\dagger \;.\label{q12}
        \eea
In view of Eqs.~(\ref{q1.2}), (\ref{q11}), and (\ref{q12}), the  Hamiltonian
takes the form:
        \be
        H=\left(\begin{array}{cc}
        D_i^\dagger D_i&0\\
        0&D_i D_i^\dagger\end{array}\right)\;. \label{q13}
        \ee
The latter relation together with Eq.~(\ref{q6}) and the identities:
        \be
        ker (D_i^\dagger D_i)=ker(D_i)\;,~~~
        ker (D_i D_i^\dagger)=ker(D_i^\dagger)\;,\label{q100}
        \ee
lead to the desired result \cite{w1}, namely
        \be
        {\rm index}_W=dim(ker\, D_i)-dim(ker\, D_i^\dagger)\;,
        \label{q14}
        \ee
for either of $i=1,2$. In fact, Witten chooses $D_1=D_2$ to  satisfy the
compatibility conditions (\ref{q9}) -- (\ref{q12}).\footnote{Note that this
is not a necessary condition for satisfying (\ref{q9}) -- (\ref{q12}).}
If now one identifies ${\cal H}_\pm$ with abstract inner product  (Hilbert)
spaces $\Gamma_1$ and $\Gamma_2$, and $D_i:\Gamma_1\to \Gamma_2$
with two (parameter dependent) Fredholm operators, then Eq.~(\ref{q14})
implies:
        \be
        {\rm index}_W=\mbox{index}^{\rm Analytic}(D_i)
        \;,
        \label{15}
        \ee
for both $i=1,2$. In particular, one can choose $\Gamma_a$ ($a=1,2$) to be
spaces of smooth sections of a pair of complex Hermitian vector bundles $E_a$
and $D_i$ a pair of elliptic differential operators. Then, one has:
        \be
        {\rm index}_W =\mbox{index}^{\rm Atiyah-Singer}(D_i)
        \;,\label{q16}
        \ee
where by the Atiyah-Singer index, we mean the {\em topological index}
introduced by Atiyah and Singer \cite{shanahan}. Eq.~(\ref{q16}) is proven for
twisted Dirac operators and other classical elliptic operators using the path
integral techniques. The former result together with a result of K-theory lead
to a proof of  the general index theorem, \cite{a-g-w,jmp94}.

\section{R-S PSQM and the Classification Problem}
Ref.~\cite{p8} presents a  detailed analysis of both the R-S and the B-D
($p=2$) PSQM. Here the relevant results are quoted without  proof for
brevity. As demonstrated in \cite{p8} the analysis of the B-D PSQM is
quite analogous to the R-S PSQM. This analogy also extends to the
subject of this article. Hence the results concerning the B-D PSQM will
not be explicitly mentioned.

Consider the R-S parasuperalgebra (\ref{q0.1}), (\ref{q0.2}) written in terms
of the self-adjoint generators (\ref{q4}):
        \bea
        Q_1^3-\{Q_1,Q_2^2\}-Q_2Q_1Q_2=0&&\label{q22}\\
        Q_2^3-\{Q_2,Q_1^2\}-Q_1Q_2Q_1=0&&\label{q23}\\
        \left[Q_1,H\right]=\left[Q_2,H\right]=0 &&\label{q24}\\
        Q_1^3=2Q_1H&&\label{q25}\\
        Q_2^3=2Q_2H&&\label{q26}\;.
        \eea
These relations are sufficient to prove the following statements \cite{p8}:
        \begin{itemize}
        \item[1)] In general, the spectrum  consists of both negative and
non-negative energy eigenvalues.
        \item[2)] The negative and zero energy eigenvalues are
non-degenerate\footnote{Here  degeneracy refers to the eigenvalues of one of
the self-adjoint charges, say $Q_1$.}.
	\item[3)] The positive energy levels may be non-degenerate, doubly degenerate
or triply degenerate. The doubly degenerate levels consist of a pair of odd and
even chirality states,
where as the triply degenerate levels involve two even (resp.\ odd) and one odd
(resp.\ even)
states.
	\item[4)] Consider an arbitrary  degenerate energy level $E$ and denote the
corresponding degeneracy subspace by ${\cal H}_E$. Then  in a basis where
$Q_1$ is diagonal, one has the following matrix representations for the
relevant operators:
	\begin{itemize}
	\item[a)] For doubly degenerate levels:
	\bea
	\left. Q_1\right|_{{\cal H}_E}= \sqrt{2E}\sigma_3\,, &&
	\left. Q_2\right|_{{\cal H}_E}= \sqrt{2E}\sigma_1\,,
	\label{rep2}\\
	\left. \tau\right|_{{\cal H}_E}=\eta\,\sigma_1\,,&&\nn
	\eea
	where $\sigma_i$ are Pauli matrices and $\eta=\pm$.
	\item[b)] For triply degenerate levels:
	\be
	\begin{array}{c}
	\left. Q_1\right|_{{\cal H}_E}\: =\: \sqrt{2E}\lll
	\begin{array}{ccc}
	1&0&0\\
	0&0&0\\
	0&0&-1
	\end{array}\rrr=\sqrt{2E}J_3^{~(1)} \;,    \\ \\
	\left. Q_2\right|_{{\cal H}_E}\: =\: \sqrt{2E}\lll
	\begin{array}{ccc}
	0&	\frac{\zeta}{\sqrt{2}}    &-i\epsilon\sqrt{1-\zeta^2}\\
	\frac{\zeta}{\sqrt{2}}&0&\frac{\zeta}{\sqrt{2}}\\
	i\epsilon\sqrt{1-\zeta^2}&\frac{\zeta}{\sqrt{2}}&0
	\end{array}\rrr \\ \\
	\hspace{1.9cm}	=\sqrt{E}\zeta J_1^{~(1)}+i\epsilon\sqrt{2E(1-\zeta^2)}\lll
	\begin{array}{ccc}
	0&0&-1\\
	0&0&0\\
	1&0&0
	\end{array}\rrr\;,	\\ \\
	\left. \tau\right|_{{\cal H}_E}=\lll
	\begin{array}{ccc}
	0&0&\tilde\eta\\
	0&\eta&0\\
	\tilde\eta&0&0
	\end{array}\rrr\;,
	\end{array}
	\label{rep}
	\ee
where $\zeta\in[0,1], ~\epsilon, ~\eta, ~\tilde\eta=\pm$ are numerical
parameters
with $\zeta\neq 0\Rightarrow \tilde\eta=-\eta$, and $J_i^{~(1)}$, with
$i=1,2,3$, are the three dimensional ($j=1$) representation of the generators
of $SU(2)$.
	\end{itemize}
	\item[5)] The non-degenerate energy eigenstates correspond to the zero
eigenvalue of $Q_1$. Indeed they are annihilated by both $Q_1$ and $Q_2$.
	\end{itemize}

In Ref.~\cite{p8}, it is argued that  in order to define an analog of the
Witten index of SQM, one must focus on PSQM systems which involve only
non-negative energy levels and triply degenerate positive energy levels. It is
also shown in \cite{p8} that postulating a particular form for the Hamiltonian,
namely Eq.~(\ref{q0.41}), originally suggested by Khare, et al \cite{kh2} for
different purposes,  one realizes the necessary conditions to define the
topological invariant $\Delta^{(p=2)}$ of Eq.~(\ref{q0.5}).  In this case, the
parameter $\zeta$ of Eqs.~(\ref{rep}) is forced to take the value $1$.

An important observation regarding this matrix representations is that
any Hamiltonian $H=H(Q_1,Q_2)=H(\qq,\qq^\dagger)$ which satisfies
(\ref{q0.1}) and (\ref{q0.2}) must necessarily respect the above matrix
representations. Therefore, these representations can be used to identify
admissible forms of the Hamiltonian. In other words, in order to classify
all the R-S PSQM Hamiltonians, one must first consider the most general
expression for $H$ and enforce the matrix representations dictated by
R-S PSQM. The only guideline for determining this general form is the fact
that the dimension of the generators $\qq$ and $\qq^\dagger$ is the 
square root of that of the Hamiltonian (energy). Therefore, in general one
must consider the following form:
	\[
	H_N=\sum_{j=1}^N [P_j(\qq,\qq^\dagger)]^{1/j}\;,\]
where $N$ is an arbitrary positive integer and $P_j(\qq,\qq^\dagger)$
is a polynomial in $\qq$ and $\qq^\dagger$ of order $2j$ with the provision
that $P_j^\dagger=P_j$. The desired classification scheme is therefore inductive
in nature. For each $N\geq 1$ one needs to write down the most general
self-adjoint polynomials $P_j$ with $j\leq N$ and then enforce the matrix
representations. This in turn leads to a series of matrix equations among
the coefficients of these polynomials. The solutions of these equations
determine the moduli space of the corresponding class of the R-S PSQM
Hamiltonians $H_N$.

In particular the classification of the Hamiltonians whose $j$-th power is
a polynomial $P_j(\qq,\qq^\dagger)$ of order $2j$ is equivalent to
solving a set of equations which are algebraic in the coefficients of $P_j$
and rational in $\zeta$ and $\epsilon\sqrt{1-\zeta^2}$. Eq.~(\ref{q0.41}) is
a particular example of such a Hamiltonian.

In the remainder of this section,  first the system of Eq.~(\ref{q0.41}) is
generalized to  systems whose Hamiltonian is  square root of a forth order
polynomial in the generators, $H=\sqrt{P_2(\qq,\qq^\dagger)}$. This leads
to a class of PSQM systems whose spectra consist only of the
non-degenerate zero energy and degenerate positive energy levels. 
Next  a classification of all such systems is carried out and  the subclass
which lacks the doubly degenerate positive energy levels is identified. 
The latter consists of  the systems for which  $\Delta^{(p=2)}$ of
Eq.~(\ref{q0.5}) is a topological invariant.

Consider, the most general self-adjoint Hamiltonian $H$ whose square is a forth
order  polynomial in the generators ${\cal Q}$ and ${\cal Q}^\dagger$ of
parasupersymmetry. Since according to (\ref{q0.1}), ${\cal Q}^3=0$, one has the
following most general form:
	\bea
	H&=&\left[
	C_1\, \qq^2\qq^{\dagger 2}+
	C_2\, \qq^{\dagger 2}\qq^2+
	C_3(\qq\qq^\dagger)^2+
	C_4(\qq^\dagger\qq)^2+\right. \nn\\
	&&C_5(\qq\qq^{\dagger 2}\qq+\qq^\dagger\qq^2\qq^\dagger)+
	C_6(\qq^2\qq^\dagger\qq+\qq^\dagger\qq\qq^{\dagger 2})+\nn \\
	&&\left. C_7(\qq\qq^\dagger\qq^2+
	\qq^{\dagger 2}\qq\qq^\dagger)\right]^{1/2}\;,
	\label{H}
	\eea
where $C_k$, $k=1,\cdots 7$ are real coefficients. In view of the defining
parasuperalgebra (\ref{q0.2}), this relation may be simplified to yield:
	\bea
	H^2&=&
	C_1\, \qq^2\qq^{\dagger 2}+
	(C_2-2C_5)\qq^{\dagger 2}\qq^2+
	C_3(\qq\qq^\dagger)^2+\nn\\
	&&(C_4-2C_5)(\qq^\dagger\qq)^2+
	\left[ 8C_5 \,\qq^\dagger\qq+4C_7(\qq^2+\qq^{\dagger 2})\right]H+ \nn\\
	&&(C_6-C_7)(\qq^2\qq^\dagger\qq+\qq^\dagger\qq\qq^{\dagger 2})
	\label{H2}
	\eea
However, one still needs to check whether this equation is compatible with
(\ref{q0.2}).

Having listed the matrix representations of the parasupersymmetry generators
for each  energy level, i.e., Eqs.~(\ref{rep2}) and (\ref{rep}), the
compatibility requirement may be enforced by substituting the matrix
representations of $\qq$, $\qq^\dagger$  and $H$ in Eq.~(\ref{H2}). This leads
to  a set of algebraic equations for the coefficients $C_k$.

Before pursuing the analysis of these equations, however, one must note that in
view of the item 5 of the above list and the form of the Hamiltonian (\ref{H}),
the non-degenerate positive energy levels are not present in the spectrum. This
is because such states,if existed, would have been annihilated by the right
hand side of Eq.~(\ref{H2}) and  survived by the left hand side, leading to an
obvious contradiction.

Next, consider the triply degenerate energy levels.  In view of the analysis of
the previous section, it is favorable to switch to a basis in which the
chirality operator $\tau$ is diagonal. It is clear from (\ref{rep}) that $\tau$
has two eigenvalues ($\pm 1$), one of them being degenerate. This allows one to
have infinitely many choices for a unitary basis which diagonalizes $\tau$. In
the following this arbitrariness is exploited to choose a basis in which the
expressions for all the operators are considerably simplified.  In fact, as it
is demonstrated instantly, the value $\zeta=0$ is forbidden by the relation
(\ref{H}). Thus $\tau$ depends only on the conventional sign $\eta$ which can
be set to $+$ without loss of generality. Here it is assumed that the choice of
$\eta$ is independent of the energy eigenvalue $E$.

In the new basis:
	\be
	e_1:=\lll
	\begin{array}{c}
	\frac{-it}{\sqrt{2}}\\
	\sqrt{1-t^2}\\
	\frac{it}{\sqrt{2}}
	\end{array}\rrr\,, ~~
	e_2:=\lll
	\begin{array}{c}
	\sqrt{\frac{1-t^2}{2}}\\
	-it\\
	-\sqrt{\frac{1-t^2}{2}}
	\end{array}\rrr\,, ~~
	e_2:=\lll
	\begin{array}{c}
	\frac{1}{\sqrt{2}}\\0\\ \frac{1}{\sqrt{2}}
	\end{array}\rrr\,,
	\label{basis}
	\ee
where $t$ is defined by:
	\be
	t:=\epsilon\sqrt{1-\zeta^2}\:\in[-1,1]\,,
	\label{t}
	\ee
one has the following  matrix representations:
	\bea
	\left. \tau \right|_{{\cal H}_E}&=&\lll
	\begin{array}{ccc}
	1&0 &0\\
	0&1&0\\
	0&0&-1
	\end{array}\rrr \,,
	\nn\\
	\left. Q_1 \right|_{{\cal H}_E}&=& \sqrt{2E}\lll
	\begin{array}{ccc}
	0&0&it\\
	0&0&\sqrt{1-t^2}\\
	-it&\sqrt{1-t^2}&0
	\end{array}\rrr \,,
	\label{Q1}\\
	\left. Q_2 \right|_{{\cal H}_E}&=&\sqrt{2E}\lll
	\begin{array}{ccc}
	0&0&1\\
	0&0&0\\
	1&0&0
	\end{array}\rrr\,.
	\label{Q2}
	\eea

Using these relations and Eq. (\ref{q4}), one then obtains the expressions for
$\qq$ and $\qq^\dagger$. The latter may be substituted in the right hand side
of Eq.~(\ref{H2}).  Equating the result with  the left hand side which is just
$E^2$ times the identity matrix, leads to four independent equations for the
eight unknowns: $C_k, ~k=1,\cdots,7$ and $t$:
	\bea
	&&(1-t^2)(1+t)C_1+(1-t^2)(1-t)C_2+(1+t)^3C_3+(1-t)^3C_4-\nn\\
	&&2(1-t^2)(1-t)C_6-2(1-t^2)(1+t)=\frac{1}{2}\;,
\label{I}\\ &&\nn\\
	&&\sqrt{1-t^2}\left[ (1-t^2)C_1-(1-t^2)C_2+(1+t)^2C_3-(1-t)^2C_4+\right.\nn\\
	&&\left.2t(1-t)C_6+2t(1+t)C_7\right]=0\;,
\label{II}\\&&\nn\\
	&&(1-t^2)\left[(1-t)C_1+(1+t)C_2+(1+t)C_3+(1-t)C_4+\right.
\nn\\
	&&\left. 2(1-t)C_6+2(1+t)C_7\right]=\frac{1}{2}\;,
\label{III}\\&&\nn\\
	&&(1-t)^2C_3+(1+t)^2C_4+2(1-t^2)C_5=\frac{1}{4}\;.
\label{IV}
	\eea
One immediately concludes from  Eq.~(\ref{III}) that the values $t=\pm1$ (i.e.,
$\zeta=0, ~\epsilon=\pm1$) are forbidden.

Eqs.~(\ref{I})--(\ref{IV}) may be solved to express four of the unknowns in
terms of the other four. For reasons which will be clear  shortly,  $t$,
$x:=C_3$, $y:=C_4$ and $z:=C_7$ are chosen as independent variables. The
solutions have a remarkably simple form:
	\bea
	C_1&=&\frac{1-4(1+t)^2x}{4(1-t^2)}\;,
\label{20.1}\\
	C_2&=&\frac{1-4(1-t)^2y}{4(1-t^2)}\;,
\label{20.2}\\
	C_5&=&\frac{1-4(1-t)^2x-4(1+t)^2y}{8(1-t^2)}\;,
\label{20.3}\\
	C_6&=&\frac{-(1+t)z}{1-t}\;.
	\label{20.4}
	\eea

At this stage, one must emphasize the role of the parameter $t$. As it is
argued in \cite{p8}, the defining parasuperalgebra does not impose any
restrictions on the value of $t$. In fact, it may depend on the energy
eigenvalue $E$. In general the value or values of $t$ may only be fixed if  the
detailed structure  of the particular system of interest is known.  For the
systems admitting a Hamiltonian of the form (\ref{H}), the coefficients $C_k$
are universal parameters independent of the energy eigenvalues.   Existence of
triply degenerate energy levels,  however, make them dependent  on  $t$. Thus
it is reasonable to assume that $t$ is also a universal (deformation) parameter
taking a single value for all the triply degenerate  energy levels.  In fact,
one can show the universality (uniqueness) of the parameter $t$ without making
any additional assumption. To see this, let us assume that there is another
parameter $t'\in(-1,1)$ associated with some other energy eigenvalue $E'>0$.
Then the same analysis applies for $E'$ and one obtains exactly the same
equations as (\ref{20.1})--(\ref{20.4}) with $t$ replaced by $t'$. Introducing:
	\[
	 \mbox{\bf${\cal C}$}:=\lll\begin{array}{c}
	C_1\\C_2\\C_5\\C_6\end{array}\rrr\;, ~~~~~
	\mbox{\bf${\cal X}$}:=\lll\begin{array}{c}
	1\\x\\y\\z\end{array}\rrr\;,\]
one may rewrite Eqs.~(\ref{20.1})--(\ref{20.4}) as a matrix equation:
	\be
	 \mbox{\bf${\cal C}$}= F(t) \mbox{\bf${\cal X}$}\;,
	\label{matrix}
	\ee
where $F(t)$ is a matrix  whose value may be easily read from
Eqs.~(\ref{20.1})--(\ref{20.4}).  Since Eq.~(\ref{matrix}) must hold for both
$t$ and $t'$, one has:
	\[  \left[ F(t)-F(t')\right]\mbox{\bf${\cal X}$}=0\;.\]
However, by definition $\mbox{\bf${\cal X}$}\neq 0$. This implies the matrix
$F(t)-F(t')$ to be singular, i.e.,
	\[ \det \left[ F(t)-F(t')\right]=0\;.\]
This equation can be easily solved. Its only solution is $t'=t$. This concludes
the proof that $t$ is independent of the energy eigenvalues.

In view of the uniqueness of $t$ and Eqs.~(\ref{20.1})--(\ref{20.4}), one may
also assert  that the moduli space ${\cal M}$ of the ($p=2$) PSQM systems
admitting a Hamiltonian of the form (\ref{H}) and possessing triply degenerate
energy levels, is $(-1,1)\times\R^3$.  ${\cal M}$ has a subspace ${\cal N}$
corresponding to systems which include doubly degenerate energy levels as well.
To construct ${\cal N}$, one may appeal to the matrix representations of the
generators of the parasupersymmetry for the doubly degenerate levels, namely
Eqs.~(\ref{rep2}). In view of these equations and relations
(\ref{20.1})--(\ref{20.4}), one can easily show that  the existence of doubly
degenerate levels fixes $x:=C_3$ and  $y:=C_4$ according to:
	\be
	x=y=\frac{1}{4}\;.
	\label{x,y}
	\ee
This can be directly inferred from the original expression for the Hamiltonian
(\ref{H}). According to Eqs.~(\ref{20.1})--(\ref{20.4}) and (\ref{x,y}),
${\cal N}$ may be identified with $(-1,1)\times\R\subset{\cal M}$. The subspace
of ${\cal M}$ including the systems whose positive energy levels are all triply
degenerate is then the set $\tilde{\cal M}:={\cal M}-{\cal N}$. If one assumes
that the chirality operator has the same representation for all the positive
energy levels, i.e., $\tilde\eta=-\eta$ of (\ref{rep}) is independent of $E$,
then the integer $\Delta^{(p=2)}$ of Eq.~(\ref{q0.5}) is clearly a topological
invariant for the elements of  $\tilde{\cal M}$.

It must also be emphasized that by the topological invariance of
$\Delta^{(p=2)}$ one means that if the Hamiltonian depends on a set of
parameters $m\in M$ , $\Delta^{(p=2)}$ is left unchanged under continuous
variations of $m$.  There is an obvious distinction between $\tilde{\cal M}$
and $M$. The latter may be an arbitrary (locally connected) topological
(parameter) space that parameterizes the operator  $\qq$ and therefore $H$. For
example, one may take $M$ to be the space of all geometries on a given
Riemannian manifold $X$, in which case $\Delta^{(p=2)}$ is a true topological
invariant of $X$. On the other hand one
may keep $\qq$ fixed and try to deform the Hamiltonian by continuously changing
the parameters $t,x,y$ and  $z$. In this case all values of the parameters
belonging to $\tilde{\cal M}$ must yield the same value for $\Delta^{(p=2)}$,
for $\tilde{\cal M}$ (with the subspace topology induced from the usual
Euclidean metric topology on ${\cal M}$) is connected.

\section{Mathematical Interpretation of Parasupersymmetric Topological
Invariants}
As it is argued in the preceding section, the topological content of the
systems under investigation is independent of the free parameters $t, ~x, ~y$
and $z$ as far as they remain in $\tilde{\cal M}$. Therefore, in general one
may fix one or some of these parameters in a topological investigation of PSQM.
In the following,  $t$ is chosen to vanish ($t=0$) while the other parameters
are kept free.   In this case,  the generators $Q_1$ and $Q_2$ have
particularly simple expressions. According to (\ref{Q1}) and (\ref{Q2}), one
has:
	\be
	\left. Q_1\right|_{{\cal H}_E}=\sqrt{2E}\left(
        \begin{array}{ccc}
        0&0&0\\
        0&0&1\\
        0&1&0\end{array}\right)\;, ~~~~~
        \left. Q_2\right|_{{\cal H}_E}=\sqrt{2E}\left(
        \begin{array}{ccc}
        0&0&1\\
        0&0&0\\
        1&0&0\end{array}\right)\;.
	\label{Qs}
	\ee

In analogy with the case of SQM, as discussed in Sec.~2, Eqs.~(\ref{Qs}) may be
employed to yield an algebraic expression for the topological invariant
$\Delta^{(p=2)}$. In order to derive such an expression, first consider the
following representation of the Hilbert space:
        \be
        {\cal H}=\left(\begin{array}{c}
        {\cal H}^1_+\\{\cal H}^2_+\\{\cal H}_-\end{array}
        \right)\;,\:\:\:{\rm with}~~~{\cal H}_+=:
        \left(\begin{array}{c}
        {\cal H}^1_+\\{\cal H}^2_+\end{array}\right)\;,
        \label{q32}
        \ee
where ${\cal H}_+$ and ${\cal H}_-$ are $+1$ and $-1$ eigenspaces of $\tau$
respectively.

In view  of the constructions (\ref{q8}) and Eqs.~(\ref{Qs}), we propose:
        \be
        Q_1=\left(\begin{array}{ccc}
        0&0&0\\
        0&0&D_1^\dagger\\
        0&D_1&0
        \end{array}\right)\;\;,\;\;
        Q_2=\left(\begin{array}{ccc}
        0&0&D_2^\dagger\\
        0&0&0\\
        D_2&0&0
        \end{array}\right)\;,
        \label{q33}
        \ee
where $D_i:{\cal H}^i_+ \to{\cal H}_-$ ($i=1,2$) are linear operators. Next we
substitute the ansatz (\ref{q33}) in the parasuperalgebra (\ref{q0.2}) or
alternatively (\ref{q22})--(\ref{q26}) and Eq.~(\ref{H}) for the Hamiltonian.

Condition ${\cal Q}^3=0$ which in terms of $Q_1$ and $Q_2$ is expressed as
Eqs.~(\ref{q22}) and (\ref{q23}), together with the ansatz (\ref{q33}) lead to
the following compatibility conditions:
 	\be
        (D_2D_2^\dagger-D_1D_1^\dagger)D_i=0~~~~
        ~~(i=1,2)\;.
        \label{q34}
        \ee
These conditions, in turn simplify the expression (\ref{H2}) for the
Hamiltonian which then reads:
	\be
	H^2=\frac{1}{2}\lll\begin{array}{ccc}
	\gamma_1(D_2^\dagger  D_2)^2&i\gamma_2D_2^\dagger  D_1D_1^\dagger  D_1&0\\
	&&\\
	-i\gamma_2D_1^\dagger  D_1D_1^\dagger  D_2&\gamma_1(D_1^\dagger  D_1)^2&0\\
	&&\\
	0&0&\gamma_3(D_1D_1^\dagger +D_2D_2^\dagger)^2+\gamma_4[(D_1D_1^\dagger)^2-
	(D_2D_2^\dagger)^2]
	\end{array}\rrr\;,
	\label{H2matrix}
	\ee
where
	\bea
	\gamma_1&:=&C_1+C_2+C_3+C_4-2C_6-2C_7\,,\nn\\
	\gamma_2&:=&C_1-C_2+C_3-C_4\,,\label{gammas}\\
	\gamma_3&:=&\frac{C_3}{2}+{C_4}{2}+C_5\,, ~~
	\gamma_4\::=\: C_6+C_7\;.\nn
	\eea
Taking $t=0$ in Eqs.~(\ref{20.1})--(\ref{20.4}) and substituting the result in
(\ref{gammas}), one finds
	\[\gamma_1=\frac{1}{2}\,, ~~ \gamma_2=\gamma_4=0\,, ~~
\gamma_3=\frac{1}{8}\;.\]
The fact that $\gamma$'s are independent of the  variables $x,y$, and $z$ is
quite remarkable. In view of these results, one can easily take the square root
of both sides of (\ref{H2matrix}) to yield:
	\be
	H=\frac{1}{2}\lll\begin{array}{ccc}
	D_2^\dagger D_2& 0&0\\
	0&D_1^\dagger D_1&0\\
	0&0&\frac{1}{2}(D_1D_1^\dagger+D_2D_2^\dagger)
	\end{array}\rrr\;,
	\label{q36}
	\ee
Another remarkable observation is that indeed the Hamiltonian as expressed by
Eq.~(\ref{q36}) also satisfies the other parasuperalgebra relations, namely
Eq.~(\ref{q0.2}) or alternatively Eqs.~(\ref{q24})--(\ref{q26}). This is also
highly nontrivial.

Having obtained the expression for the Hamiltonian in a basis which explicitly
distinguishes the odd and even chirality states, one can easily derive the
formula for $\Delta^{(p=2)}$:
        \be
        \ddd=dim(ker~D_1)+dim(ker~D_2)-2dim(ker~D_1^\dagger
        \cap ker~D_2^\dagger)\;.
        \label{q39}
        \ee
Here we have employed the following identifications:
        \bea
        n_0^{\pi B}&=&dim(ker~D_1^\dagger D_1\oplus ker~D_2^\dagger D_2)
        \nn\\
                   &=&dim(ker~D_1)+dim(ker~D_2)
	\label{q37}\\
        n_0^{\pi F}&=&dim (ker~[D_1\,D_1^\dagger+D_2\,D_2^\dagger])\nn\\
        &=&dim(ker~D_1^\dagger \cap ker~D_2^\dagger)\;.
        \label{q38}
        \eea
In Eqs.~(\ref{q37}) and (\ref{q38}) use is made of relations~(\ref{q100}).

It turns out that conditions (\ref{q34}) may be used to simplify the expression
(\ref{q39}) for $\ddd$. To see this let us define $A_i:=D_iD_i^\dagger$,
(i=1,2). Then multiplying Eqs.~(\ref{q34}) by $D_i^\dagger$ from the right and
writing the resulting equations in terms of $A_i$, one has:
        \be
        (A_1-A_2)A_1=0~,~~~~(A_1-A_2)A_2=0\;.
        \label{q120}
        \ee
In view of the fact that $A_i$ are self-adjoint and positive (semi)definite
operators, Eqs.~(\ref{q120}) imply $ker~A_1=ker~A_2$. This together with the
identities~(\ref{q100}) lead to $ker~D_1^\dagger=ker~D_2^\dagger$.
Thus, we have:
        \be
        \ddd={\rm index}^{\rm analytic}(D_1)+
        {\rm index}^{\rm analytic}(D_2)\;.
        \label{q121}
        \ee
Eq.~(\ref{q121}) provides the desired mathematical interpretation for the
parasupersymmetric topological invariant (\ref{q0.5}). 

It must be emphasized that Eq.~(\ref{q121}) is only valid for the ansatz
(\ref{q33}).  In fact the most general expression which relates
$Q_1$ and $Q_2$ with a pair of linear operators $D_1$ and $D_2$ is
(\ref{q8}), where the representation (\ref{q7}) is used for the Hilbert
space. In this case however, enforcing the R-S PSQM
algebra, one is  led to complicated compatibility conditions between
$D_1$ and $D_2$ which render a similar approach ineffective.

\section{Conclusion}
The ($p=2$) parasupersymmetric quantum mechanics may be viewed as a
generalization of the ordinary supersymmetric quantum mechanics. A study of the
spectrum degeneracy structure of the ($p=2$) parasupersymmetry leads to the
definition of a topological invariant. For a class of ($p=2$)
parasupersymmetric systems this invariant may be given a well-known
mathematical meaning, namely that it is associated with the sum of analytical
indices of a pair of Fredholm operators. In a sense, this is a negative result
as one might have hoped for a more general and possibly new topological
invariant.

Unlike supersymmetric quantum mechanics, the form of the Hamiltonian is not
determined by the defining algebraic relations in parasupersymmetric quantum
mechanics.  Thus in general one needs to investigate possible forms of the
Hamiltonian which are compatible with the defining parasuperalgebras of
($p=2$)--PSQM and attempt to classify the corresponding systems. In the present
article it is shown how the matrix representations of the relevant operators,
which are valid for any quantum system satisfying the definition of PSQM, may
be used to classify the forms of the corresponding Hamiltonians. This is
carried out explicitly for the case where the Hamiltonian is the square root of
a forth order polynomial in the generators. However the method enjoys general
applicability. 

For the particular class (\ref{H}) of the Hamiltonians considered here,
it is shown that the parameter $t=\epsilon\sqrt{1-\zeta^2}$ of the R-S PSQM
is a universal parameter, i.e., it is independent of the energy eigenvalues.

The mathematical interpretation of parasupersymmetric topological invariant
$\ddd$ offered in this article depends on the ansatz (\ref{q33}) chosen to
relate the generators of parasupersymmetry with some Fredholm operators. This
is justified by making analogy with the case of supersymmetry and assuming  the
parameter $t$ to vanish. In fact, if one does retain the ansatz (\ref{q33}) but
considers the case $t\neq 0$, then the parasuperalgebra relations are not as
trivially satisfied. In fact, in this case, Eq.~(\ref{H2matrix}) remains valid
but the compatibility  with Eq.~(\ref{q0.2}) leads to:
	\be
	\gamma_1-\gamma_2=\frac{1}{2}\,, ~~\gamma_3=\frac{1}{8}\;.
	\label{qg}
	\ee
These equations are obtained by multiplying both sides of Eq.~(\ref{q0.2}) by
$H$ and using the same equation to express the left hand side of the result in
terms of $\qq$ and $\qq^\dagger$. This yields an equation involving $H^2$,
$\qq$ and $\qq^\dagger$ which upon substitution of  (\ref{q4}), (\ref{q33}) and
(\ref{H2matrix}) results in (\ref{qg}). In view of the definition of $\gamma$'s
(\ref{gammas}), the latter  equations add to Eqs.~(\ref{I})--(\ref{IV}). This
indicates that  the ansatz (\ref{q33}) is valid for a proper subset of
$\tilde{\cal M}$ consisting of the sector corresponding to $t=0$ and one
defined by the simultaneous solution of Eqs.~(\ref{I})--(\ref{IV}) and
(\ref{qg}) with $t\neq 0$.

The methods and ideas developed in this article may be easily applied
for the B-D PSQM. In fact, requiring the Hamiltonian to have the form
(\ref{H}) and using the matrix representations of the generators of
the B-D PSQM \cite{p8}, one recovers the same equations for
the coefficients, i.e., (\ref{20.1})--(\ref{20.4}), with $t=0$. Therefore
the results obtained in sections~3 and~4 for the R-S PSQM with $t=0$
are also valid for  the B-D PSQM systems in general.

\section*{Acknowledgements}
Parts of this research were carried out at the Institute for Studies in
Theoretical Physics and Mathematics (IPM) at Tehran. The author would like to
thank V.\ Karimipour, S.\ Rouhani of the IPM, and A.\ Rezaii of Tabriz
University for invaluable comments and discussions. He also appreciates the
financial support of the Killam foundation of Canada.

\end{document}